\documentclass[]{article}

\usepackage{mysprocl}
\usepackage{cite}
\usepackage{epsfig}
\catcode`@=11
\def\citer{\@ifnextchar [{\@tempswatrue\@citexr}{\@tempswafalse\@citexr[]}}
 
\def\@citexr[#1]#2{\if@filesw\immediate\write\@auxout{\string\citation{#2}}\fi
  \def\@citea{}\@cite{\@for\@citeb:=#2\do
    {\@citea\def\@citea{--\penalty\@m}\@ifundefined
       {b@\@citeb}{{\bf ?}\@warning
       {Citation `\@citeb' on page \thepage \space undefined}}%
\hbox{\csname b@\@citeb\endcsname}}}{#1}}
\catcode`@=12

\def\reffi#1{\mbox{Fig.~\ref{#1}}}

\def\refta#1{\mbox{Tab.~\ref{#1}}}
\def\citere#1{\mbox{Ref.~\cite{#1}}}

\newcommand{\mste}{m_{\tilde{t}_1}}
\newcommand{\mstz}{m_{\tilde{t}_2}}

\newcommand{\At}{A_t}

\newcommand{\Xt}{X_t}

\newcommand{\msusy}{M_{\mathrm{SUSY}}}

\newcommand{\cp}{{\cal CP}}
\newcommand{\wz}{\sqrt{2}}

\def\ed#1{\frac{1}{#1}}
\newcommand{\twol}{two-loop}
\newcommand{\onel}{one-loop}

\newcommand{\MW}{M_W}
\newcommand{\MZ}{M_Z}

\newcommand{\mh}{m_h}

\newcommand{\mt}{m_{t}}

\newcommand{\Stop}{\tilde{t}}

\newcommand{\tsf}{\theta\kern-.20em_{\tilde{f}}}
\newcommand{\tsfp}{\theta\kern-.20em_{\tilde{f}\prime}}
\newcommand{\tsq}{\theta\kern-.15em_{\tilde{q}}}

\newcommand{\sweff}{\sin^2\theta_{\mathrm{eff}}}

\newcommand{\KL}{\left(}
\newcommand{\KR}{\right)}

\newcommand{\VL}{\left( \begin{array}{c}}
\newcommand{\VR}{\end{array} \right)}
\newcommand{\ML}{\left( \begin{array}{cc}}
\newcommand{\MLd}{\left( \begin{array}{ccc}}
\newcommand{\MLv}{\left( \begin{array}{cccc}}
\newcommand{\MR}{\end{array} \right)}

\newcommand{\CTb}{\cot \beta\hspace{1mm}}

\newcommand{\gev}{\,\, \mathrm{GeV}}
\newcommand{\mev}{\,\, \mathrm{MeV}}

\newcommand{\BC}{\begin{center}}
\newcommand{\EC}{\end{center}}
\newcommand{\BE}{\begin{equation}}
\newcommand{\EE}{\end{equation}}
\newcommand{\BEA}{\begin{eqnarray}}
\newcommand{\BEAnn}{\begin{eqnarray*}}
\newcommand{\EEA}{\end{eqnarray}}
\newcommand{\EEAnn}{\end{eqnarray*}}

\newcommand{\id}{{\rm 1\kern-.12em
\rule{0.3pt}{1.5ex}\raisebox{0.0ex}{\rule{0.1em}{0.3pt}}}}

\def\al{\alpha}

\def\de{\delta}

\def\De{\Delta}

\newcommand{\lc}{linear collider}
\newcommand{\epem}{$e^+e^-$}

\newcommand{\gvf}{g_V^f}
\newcommand{\gaf}{g_A^f}

\begin{document}

\thispagestyle{empty}
\setcounter{page}{0}
\def\thefootnote{\fnsymbol{footnote}}

\begin{flushright}
DESY 99--117\\
KA--TP--14--1999\\
TTP99--37\\
hep-ph/9909538 \\
\end{flushright}

\vspace{1cm}

\begin{center}

{\large\sc {\bf Implications of Results }}

\vspace*{0.4cm} 

{\large\sc {\bf from $Z$- and $WW$-Threshold Running}}
\footnote{talk given by S.~Heinemeyer at the {\em International
    Workshop on Linear Colliders},\\
 \mbox{}\hspace{2em} April 28th - May 5th 1999, Sitges, Spain}

\vspace{1cm}

{\sc S.~Heinemeyer$^{1\,}$%
\footnote{
email: Sven.Heinemeyer@desy.de
}%
, Th.~Mannel$^{2\,}$%
\footnote{
email: Thomas.Mannel@physik.uni-karlsruhe.de 
}%
 and G.~Weiglein$^{3\,}$%
\footnote{
email: Georg.Weiglein@cern.ch
}%
}

\vspace*{1cm}

$^1$ DESY Theorie, Notkestr. 85, D--22603 Hamburg, Germany

\vspace*{0.4cm}

$^2$ Institut f\"ur theoretische Teilchenphysik, 
     Universit\"at Karlsruhe,\\ D--76128 Karlsruhe, Germany

\vspace*{0.4cm}

$^3$ Institut f\"ur theoretische Physik, 
     Universit\"at Karlsruhe,\\ D--76128 Karlsruhe, Germany \\
     CERN, TH Division, CH--1211 Geneva 23, Switzerland

\end{center}

\vspace*{1cm}

\begin{abstract}
One year of $Z$- and $WW$-Threshold running of TESLA can provide the
possibility to measure electroweak precision observables to an 
extremely high
accuracy. At the $Z$~peak ${\cal O}(10^9)$ $Z$~bosons and about $6
\times 10^8$ $b$~quarks can be collected. 
We employ the expected uncertainties 
$\De\MW = 6 \mev$ and $\De\sweff = 0.00001$ and demonstrate in this way
that very stringent consistency tests of the Standard Model (SM) and the 
Minimal Supersymmetric Standard Model (MSSM) will be possible. The 
indirect determination of the Higgs-boson mass within the SM can reach 
an accuracy of about $5 \%$.
The $6 \times 10^8$ $b$ quarks can be used to investigate various $b$
physics topics.

\end{abstract}

\def\thefootnote{\arabic{footnote}}
\setcounter{footnote}{0}

\newpage



\title{Implications of Results from $Z$- and $WW$-Threshold Running}

\author{S.~HEINEMEYER}
\address{DESY Theorie, Notkestr. 85, D--22603 Hamburg, Germany}
\author{$\qquad\qquad\qquad\qquad$ TH.~MANNEL \hfill 
        G.~WEIGLEIN $\qquad\qquad\qquad \qquad$}
\address{$\qquad$~ Institut f\"ur Theoretische Teilchenphysik \hfill
         Institut f\"ur Theoretische Physik $\qquad$~\mbox{} \\
         Universit\"at Karlsruhe, D--76128 Karlsruhe, Germany}
\maketitle
\abstracts{
One year of $Z$- and $WW$-Threshold running of TESLA can provide the
possibility to measure electroweak precision observables to an 
extremely high
accuracy. At the $Z$~peak ${\cal O}(10^9)$ $Z$~bosons and about $6
\times 10^8$ $b$~quarks can be collected. 
We employ the expected uncertainties 
$\De\MW = 6 \mev$ and $\De\sweff = 0.00001$ and demonstrate in this way
that very stringent consistency tests of the Standard Model (SM) and the 
Minimal Supersymmetric Standard Model (MSSM) will be possible. The 
indirect determination of the Higgs-boson mass within the SM can reach 
an accuracy of about $5 \%$.
The $6 \times 10^8$ $b$ quarks can be used to investigate various $b$
physics topics.
}


\section{Theoretical basis}

Electroweak precision observables (POs) provide an excellent tool
to either distinguish different models from each other or to perform
internal consistency tests of a specific model.
LEP2, SLD and the Tevatron have reached an accuracy of 
$\De\MW = 42 \mev$ for the $W$-boson mass and $\De\sweff = 0.00018$
for the effective leptonic mixing angle~\cite{lep2sldtevatron}.
$\De\MW$ will be improved by new data from these experiments and by
data from the future LHC.
A future \epem\ \lc\ (LC) might be able to reach an even better accuracy for
the POs. 
At TESLA it is planned to realize a high luminosity of 
${\cal L} \approx 7 \cdot 10^{33} {\rm cm}^{-2} {\rm s}^{-1}$ at low
energies. 
In a one-year run of TESLA at the 
$2 \MW$~threshold $\MW$ can be pinned down to $\De\MW~=~6\mev$. A
one-year run at the $Z$ pole would provide ${\cal O}(10^9)$ $Z$~bosons
and could thus determine the effective leptonic mixing angle up to
$\De\sweff = 0.00001$~\cite{gigazacc}. 
This low-energy run scenario with high luminosity we will refer to as
`GigaZ' in the following.
In \refta{tab:precallcoll} we show the expected experimental accuracies
for $\MW$, $\sweff$, the top-quark mass $\mt$, and the (lightest) Higgs-boson 
mass $\mh$ in this scenario in comparison to LEP2/Tevatron, the LHC, and
a LC with lower luminosity.%
\footnote{
In the case of the SM $\mh$ denotes the Higgs-boson mass, in the case
of the MSSM $\mh$ denotes the mass of the lightest $\cp$-even Higgs
boson.}
%
\begin{table}[ht!]
\renewcommand{\arraystretch}{1.5}
\BC
\begin{tabular}{|c||c|c|c|c|}
\cline{2-5} \multicolumn{1}{c|}{}
  & LEP2/Tevatron & LHC & LC & GigaZ \\ \hline \hline
$\MW$     & 30 MeV  & 15 MeV  & 15 MeV   & 6 MeV      \\ \hline
$\sweff$  & 0.00018 & 0.00018 & 0.00018  & 0.00001    \\ \hline\hline
$\mt$     & 4 GeV   & 2 GeV   & 0.2 GeV  & 0.2 GeV    \\ \hline
$\mh$     & ?       & 0.2 GeV & 0.05 GeV & 0.05 GeV   \\ \hline
\end{tabular}
\EC
\renewcommand{\arraystretch}{1}
\caption[]{Expected precisions of todays and (possible) future
  accelerators for $\MW$, $\sweff$, $\mt$, and the (lightest) Higgs-boson
  mass $\mh$. 
  }
\label{tab:precallcoll}
\end{table}

In this paper we compare the theoretical predictions for $\MW$ and
$\sweff$ in different scenarios with the expected experimental
uncertainties. We also investigate the indirect
determination of the Higgs-boson mass in the SM.
In addition, during the one-year run at the $Z$ peak, about 
$6 \times 10^8$ $b$ quarks can be collected, allowing us to 
explore the
potential for constraining the unitarity triangle and rare $b$ decays.


\smallskip
In order to calculate the $W$-boson mass in the SM and the MSSM we use
\BE
\MW = \frac{\MZ}{\wz} \sqrt{1 + \sqrt{\frac{4\pi\al}{\wz G_F \MZ^2}
                                      \frac{1}{1 - \De r}}}~,
\label{mwdeltar}
\EE
where the loop corrections are summarized in $\De r$~\cite{delr}.
The quantity $\sweff$ is defined through the
effective couplings $\gvf$ and $\gaf$ of the $Z$ boson to fermions:
\BE
\sweff = \ed{4\,Q_f} 
  \KL 1 - \frac{\mbox{Re}\, \gvf}
               {\mbox{Re}\, \gaf} \KR~,
\EE
where the loop corrections 
are contained in $g^f_{V,A}$.
In our analysis we include the complete one-loop results for
$\MW$ and $\sweff$ in the SM and the MSSM~\cite{delr,mssm1l}
as well as the leading higher-order QCD and electroweak 
corrections~\cite{sm2lqcd,sm3lqcd,sm2lmt4,mssm2lqcd}. In order to allow
a direct comparison between the virtual effects within the SM and the
MSSM, we do not include the recent electroweak two-loop results in the
SM~\cite{sm2lnl}, for which so far no counterpart exists in the MSSM. 
In the following we will neglect the theoretical uncertainties due to
unknown higher-order corrections, but will concentrate on the impact of
an improved accuracy of both the observables $\MW$ and $\sweff$ as well
as of the input parameters $\mt$, $\mh$, etc.\ entering the theoretical
predictions.


\smallskip
In the SM the Higgs-boson mass is a free parameter. Contrary to this,
in the MSSM the masses of the neutral $\cp$-even Higgs bosons are
calculable in terms of the other MSSM parameters. The largest
corrections arise from the $t$--$\Stop$-sector, where the dominant
contribution reads:
\BE
\De \mh^2 \sim \frac{\mt^4}{\MW^2}
                \log \KL \frac{\mste^2\,\mstz^2}{\mt^4} \KR 
        \approx \frac{\mt^4}{\MW^2} \log \KL 
\frac{(\msusy^2 + \mt^2)^2 - \mt^2 \Xt^2}{\mt^4} \KR~.
\EE
Here $\mt\,\Xt = \mt (\At - \mu \CTb)$ denotes the non-diagonal entry
in the $\Stop$-mass matrix. 
Since the \onel\ corrections are known to be very large, we use the
currently most precise \twol\ result based on explicit
Feynman-diagrammatic calculations~\cite{mhiggsletterlong,mhiggslle},
where the numerical evaluation is based on~\citere{feynhiggs}.


\section{Numerical analysis of different scenarios}

\begin{figure}[ht!]
\BC
\psfig{figure=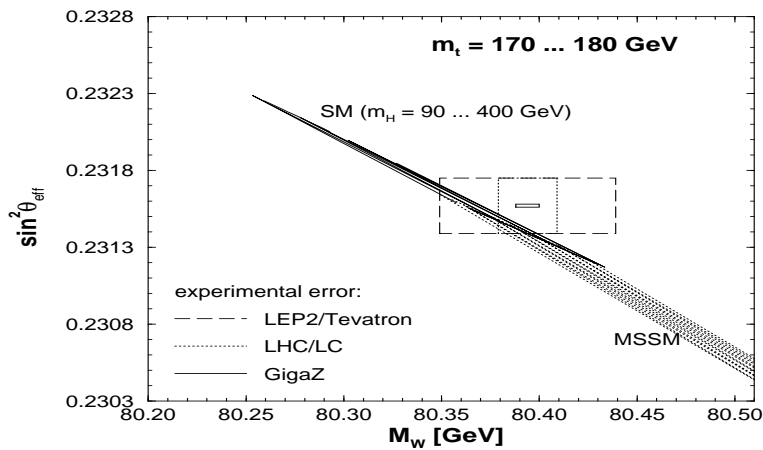,height=6cm,width=10cm}
\EC
\caption{Theoretical prediction of the SM and the MSSM in the 
$\sweff$--$\MW$-plane compared with expected experimental accuracies at
LEP2/Tevatron, the LHC and GigaZ.
\label{fig:swmw}}
\end{figure}

In order to visualize the potential of the improved accuracy obtainable at
GigaZ, we show in \reffi{fig:swmw} the
regions in the $\sweff$--$\MW$-plane which are allowed in the SM and the
MSSM. They are compared with the experimental values, taking into
account todays and possible future accuracies. The allowed region of the
SM prediction corresponds to varying $\mh$ in the interval $90 \gev \leq
\mh \leq 400 \gev$ and $\mt$ within its experimental uncertainty, while 
in the region of the MSSM prediction besides the uncertainty of $\mt$ also
the SUSY parameters are varied. As can be seen in the figure, GigaZ will
provide a very sensitive test of the theory via the precision
observables. If the experimental values of the latter should stay within
the present 1--$\sigma$ bounds it will be difficult to distinguish
between the SM and the MSSM from the precision data, but there will be
a high sensitivity to deviations from both models.

A possible obstacle for future PO analyses is the uncertainty in
$\De\al$. While 
todays most precise theory-driven 
determination ($\de\De\al = 0.00017$)~\cite{delalphatheorydriven}
could still be a limiting factor for the precision tests, it can be
shown~\cite{GigaZlong} that this would no longer be the case with an 
optimistic future expectation of $\de\De\al =
0.000075$~\cite{delalphajegerlehner}. 

\begin{figure}[ht!]
\BC
\psfig{figure=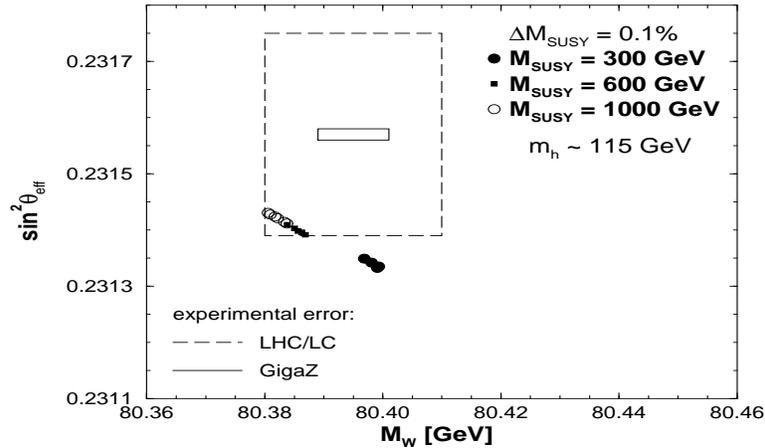,height=6cm,width=10cm}
\EC
\caption{The expected accuracy of the LHC and GigaZ is compared with the
  prediction of the MSSM for $\mh \approx 115 \gev$ and
  $\msusy = 300, 600, 1000 \gev$.
\label{fig:susy}}
\end{figure}

In the following scenario we assume that SUSY will have been
discovered and can be measured at a LC, resulting in a precision of
$\De\msusy \approx 0.1\%$. 
We also assume that the lightest
Higgs boson will have been discovered, where we take (as an example)
$115 \gev \leq \mh \leq 116 \gev$.
In \reffi{fig:susy} we compare the expected accuracies of the LHC and
GigaZ with the prediction of the MSSM for $\MW$ and $\sweff$. 
The high resolution of GigaZ is shown to have a drastic effect on the
sensitivity with which the model can be probed.


\section{Indirect determination of $\mh$}

Assuming the SM to be valid, we have performed an indirect determination
of the Higgs-boson mass separately from the predictions for $\MW$ and
for $\sweff$. Here we have made use of the numerical formulas
given in \citere{dgps}. As shown in \refta{tab:indirectmh}, the accuracy
at GigaZ would lead to a large improvement of the indirect limits on
$\mh$. In the most optimistic case, where we have assumed $\de\De\al =
0.000075$~\cite{delalphajegerlehner} and $\De\mt = 0.2$~GeV, 
$\mh$ can be predicted up to an
accuracy of about $5 \%$, which would provide an extremely sensitive test of
the SM.

\begin{table}[htb!]
\renewcommand{\arraystretch}{1.5}
\BC
%
\begin{tabular}{|l||c|c|c|c|c|}

\cline{2-6} \multicolumn{1}{c|}{}
         & LEP2/Tevatron & LHC  & LC   & GigaZ & GigaZ (most opt. case) \\ 
                                                       \hline \hline
$\MW$    & 63\%          & 33\% & 30\% & 16\%  & 12\% \\ \hline
$\sweff$ & 39\%          & 38\% & 37\% & 13\%  &  6\% \\ \hline
\end{tabular}
\EC
\renewcommand{\arraystretch}{1}
\caption[]{Expected precisions of indirect determinations of the
  Higgs-boson mass $\De\mh/\mh$ via $\MW$ and $\sweff$ for the expected 
  uncertainties for these observables at different accelerators.}
\label{tab:indirectmh}
\end{table}


\section{$b$ Quark Physics with $2 \times 10^9$ $Z$ Bosons}
GigaZ offers interesting possibilities for $b$ quark physics.
Assuming a sample of $2 \times 10^9$ $Z$ bosons produced per year 
at GigaZ, one ends up with $6 \times 10^8$ $b$ 
or $\bar{b}$ quarks. To evaluate the $b$-physics potential of GigaZ,
one has to compare it to the second generation $b$ 
physics experiments such as LHC-b or BTeV. 

As a first step towards a comparison we collect a few parameters.
The number of $b$ quarks produced per year at a second generation 
hadronic fixed target experiment such as LHC-b or BTeV is typically
$10^{12}$ to $10^{13}$ \cite{LHC-bTDR}. However,  this large
number is contained in an enormous background, the signal-to-noise
ratio being typically $S/N \approx 5 \times 10^{-3}$, to be
compared to $S/N \approx 0.15$ at GigaZ.
 
Another advantage of GigaZ over LHC-b / BTeV is the efficiency of 
the flavor tag, i.e.\ to discriminate between $b$ and $\bar{b}$. This 
tag is essential for performing $\cp$ violation studies. Using polarized
beams at GigaZ will result in a large forward-backward asymmetry
between $b$ and $\bar{b}$ quarks which -- together with the cleaner 
environment -- will allow for an efficiency of 
the flavor tag of about 60\% at GigaZ. The  
corresponding number at the hadronic fixed target experiments is 
about 6\%.  

In both GigaZ as well as LHC-b / BTeV all 
bottom flavored hadrons are accessible. In order to estimate the 
production rate for each species one may use the $b$ quark branching ratios 
obtained from LEP or from theoretical estimates \cite{LEPHFWG}. These are 
listed in~\refta{tab1}.

\begin{table}[h]
\renewcommand{\arraystretch}{1.2}
\BC
\begin{tabular}{|l|c||l|c|}
\hline
mode & $b$ branching ratio & mode & $b$ branching ratio \\
\hline
$b \to B_u$ & 40\% & $b \to B_d$ & 40\% \\
\hline
$b \to B_s$ & 12\% & $b \to \Lambda_b$ & 8\% \\
\hline
$b \to B^{**}$ & $\approx$ 25\% & $b \to B_c$ & $\approx 10^{-3} - 10^{-4}$ \\
\hline
$b \to (bcq)$-baryon & $\sim 10^{-5}$ & & \\
\hline
\end{tabular}
\EC
\renewcommand{\arraystretch}{1}
\caption{Rule-of-thumb numbers for $b$ quark branching fractions.}
\label{tab1}
\end{table}

One of the most important physics topics 
is the precision measurement of $\cp$ 
violation in $B$ decays, assuming that the CKM sector of the 
SM survives the test at the first generation $B$ 
physics experiments. The second generation will provide precise 
measurements of the CKM angles $\alpha$, $\beta$, $\gamma$ and of 
the $B_s$ mixing phase $\delta \gamma$. 

The ``gold-plated way'' to access the angle $\beta$ is to measure 
the time dependent $\cp$ asymmetry in  $B \to J/\Psi K_s$, 
which is related to $\beta$ with practically no hadronic uncertainty.
While at the first generation experiments the uncertainty
$\sigma(2\beta)$ will be about 8\%, the precision achieved at LHC-b
is expected to be $\sigma(\sin 2\beta)\approx 1.5\%$ \cite{LHC-bTDR}.
This has to be compared to $\sigma(\sin 2\beta)\approx 4\%$ at GigaZ
\cite{Hawkings}, 
thus the large statistics of LHC-b wins by a small margin.

The extraction of the CKM angle $\alpha$ is more severely affected by 
hadronic uncertainties. One way, which has been studied in some detail, 
is to measure the $\cp$ asymmetries in the decays $B \to \pi \pi$, 
from which $\alpha$ can be obtained through an isospin analysis.
This, however, requires a measurement of the decay 
$B_d \to \pi^0 \pi^0$, which is estimated to have a small branching ratio
and is very hard to identify at LHC-b / BTeV. The prospects of
extracting $\alpha$ at LHC-b in this way have been studied and yield 
an uncertainty $\sigma(\alpha) \sim 3^\circ - 10^\circ$, 
depending on the value of $\alpha$. The crucial point is whether 
a measurement of $B_d \to \pi^0 \pi^0$ will be possible at GigaZ, but
detailed studies have not yet been performed. 

Determining $\gamma$ at the first generation experiments will be extremely 
difficult, since the $B_s$ states are not accessible. At LHC-b one can 
determine $\gamma$ as well as $\delta \gamma$ from $B_s$ decays, the typical 
uncertainties being $\sigma(\gamma) \sim 6^\circ - 14^\circ$ depending on 
the $B_s$ mixing parameter and strong phases. The relative angle 
$\delta \gamma$ can be determined from a polarization analysis of the decay
$B_s \to J/\Psi \phi$ from which one expects typically 
$\sigma(\delta\gamma) \sim 10^{-2}$. However, such a measurement will
be difficult at GigaZ. Other modes relevant for $\gamma$ such as
$B \to D K$ modes have not yet been studied in detail. 

Other $b$ physics topics which can be covered 
by GigaZ are rare decays, heavy hadron spectroscopy (such as doubly
heavy hadrons), and studies taking advantage of the large polarization
of the $b$ quarks from $Z$ decays. None of these has been studied in 
detail yet, but clearly the large statistics of LHC-b / BTeV will be
hard to beat, at least in the modes which are easy to detect. 
Hence GigaZ can only have a chance using decay modes where 
a particularly clean environment is mandatory.



\bigskip
\section*{Acknowledgments}
We thank R.~Hawkings, W.~Hollik, K.~M\"onig and P.~Zerwas for helpful
discussions.





\begin{thebibliography}{00}  

\bibitem{gigazacc} K.~M\"onig, 
                   {\em these proceedings}.

\bibitem{lep2sldtevatron} M.W.~Krasny, 
             {\em Summary talk of the XXXIVth Rencontres de Moriond}

\bibitem{delr} A.~Sirlin, 
               {\em Phys. Rev.} {\bf D 22} (1980) 971.

\bibitem{mssm1l} 
               P.~Chankowski, A.~Dabelstein, W.~Hollik,
               W.~M\"osle, S.~Pokorski and J.~Rosiek,
               {\em Nucl. Phys.} {\bf B 417} (1994) 101;
               D.~Garcia, J.~Sol\`a,
               {\em Mod. Phys. Lett.} {\bf A 9} (1994) 211;
               D.~Garcia, R.~Jim\'enez, J.~Sol\`a,
               {\em Phys. Lett.} {\bf B 347} (1995) 309;
               A.~Dabelstein, W.~Hollik and W.~M\"osle,
               hep-ph/9506251;
               P.~Chankowski and S.~Pokorski,
               {\em Nucl. Phys.} {\bf B 475} (1996)~3;
               W.~de~Boer, A.~Dabelstein, W.~Hollik, W.~M\"osle and
               U.~Schwickerath,
               {\em Z. Phys.} {\bf C 75} (1997)~625.

\bibitem{sm2lqcd} A.~Djouadi and C.~Verzegnassi,
                  {\em Phys. Lett.} {\bf B 195} (1987) 265;
                  A.~Djouadi,
                  {\em Nuovo Cim.} {\bf A 100} (1988) 357;
                  B.~Kniehl, {\em Nucl. Phys.} {\bf B 347} (1990) 89.

\bibitem{sm3lqcd} K.~Chetyrkin, J.~H.~K\"uhn and M.~Steinhauser,
                  {\em Phys. Lett.} {\bf B 351} (1995) 331;
                  L.~Avdeev, J.~Fleischer, S.~Mikhailov and O.~Tarasov,
                  {\em Phys. Lett.} {\bf B 336} (1994) 560,
                  E: {\em Phys. Lett.} {\bf B 349} (1995) 597.

\bibitem{sm2lmt4} 
              J.~van der Bij and F.~Hoogeveen,
              {\em Nucl. Phys.} {\bf B 283} (1987) 477;
              R.~Barbieri, M.~Beccaria, P.~Ciafaloni, G.~Curci and A.~Vicere,
              {\em Phys. Lett.} {\bf B 288} (1992) 95;
              {\em erratum:} {\bf B 312} (1993) 511;
              {\em Nucl. Phys.} {\bf B 409} (1993) 105;
              J.~Fleischer, O.V.~Tarasov and F.~Jegerlehner,
              {\em Phys. Lett.} {\bf B 319} (1993) 249.

\bibitem{mssm2lqcd} A.~Djouadi, P.~Gambino, S.~Heinemeyer, W.~Hollik,
                    C.~J\"unger and G.~Weiglein,
                    {\em Phys. Rev. Lett.} {\bf 78} (1997) 3626;
                    {\em Phys. Rev.} {\bf D 57} (1998) 4179.

\bibitem{sm2lnl} G.~Degrassi, P.~Gambino and A.~Vicini, 
                  {\em Phys. Lett.} {\bf B 383} (1996) 219;
                  G.~Degrassi, P.~Gambino and A.~Sirlin,
                  {\em Phys. Lett.} {\bf B 394} (1997) 188;
                  S.~Bauberger and G.~Weiglein, 
                  {\em Phys. Lett.} {\bf B 419} (1998) 333.


\bibitem{mhiggsletterlong} S.~Heinemeyer, W.~Hollik and G.~Weiglein,
                    {\em Phys. Rev.} {\bf D 58} (1998) 091701;
                    {\em Phys. Lett.} {\bf B 440} (1998) 296;
                    {\em Eur. Phys. Jour.} {\bf C 9} (1999) 343. 

\bibitem{mhiggslle} S.~Heinemeyer, W.~Hollik and G.~Weiglein,
                    {\em Phys. Lett.} {\bf B 455} (1999) 179.

\bibitem{feynhiggs} S.~Heinemeyer, W.~Hollik and G.~Weiglein,
                    to appear in {\em Comp. Phys. Comm.}, 
                    hep-ph/9812320.

\bibitem{delalphatheorydriven}
                   J.~K\"uhn, M.~Steinhauser,
                   {\em Phys. Lett.} {\bf B 437} (1998) 425;
                   M.~Davier, A.~H\"ocker,
                   {\em Phys. Lett.} {\bf B 435} (1998) 427.

\bibitem{GigaZlong} S.~Heinemeyer et al, 
                    {\em in preparation}.

\bibitem{delalphajegerlehner}
               F.~Jegerlehner, talk given at the LNF spring school 1999.    

\bibitem{dgps} G.~Degrassi, P.~Gambino, M.~Passera and A.~Sirlin,
               {\em Phys. Lett.} {\bf B 418} (1998) 209.

\bibitem{LHC-bTDR} {\it LHC-b Technical design report} available under
                   http://lhcb.cern.ch/ 

\bibitem{LEPHFWG} {\it Review of Particle Properties 1998},
                  {\em Eur. Phys. Jour.} {\bf C 3} (1998); 
                  K~ Kolodziej and R.~R\"uckl, 
                  {\em Nucl. Instrum. Meth.} {\bf A 408} (1998) 33.
\bibitem{Hawkings} R. Hawkings, talk at the ECFA/DESY Linear Collider
                   Workshop, Oxford, 20-23 March 1999.  






\end{thebibliography}
\end{document}